\documentclass[epsf,useAMS,usenatbib,usegraphicx]{mn2e} 
\usepackage{epsf} \usepackage[usenames]{color}

\title[A 30-kG magnetic field in HD\,75049]{A rival for Babcock's star: the 
extreme 30-kG variable magnetic field in the Ap star HD\,75049\thanks{Based on 
observations collected at the European Southern Observatory, Chile, in the 
programme 080.D-0170(A) and as part of programmes 078.D-0080(A) and 078.D-0192(A).}}

\author[Elkin et al.] {V.G.~Elkin$^{1}$, 
G.~Mathys$^{2}$, D.W.~Kurtz$^{1}$, S. Hubrig$^{3}$, L.M.~Freyhammer$^{1}$ \\ 
$^{1}$Jeremiah Horrocks Institute of Astrophysics, University 
of Central Lancashire, Preston PR1 2HE, UK\\ 
$^{2}$European Southern Observatory, Casilla 19001, Santiago 19, Chile \\ 
$^{3}$Astrophysical Institute Potsdam, D-14482, Potsdam, Germany \\} 

\begin{document} 
\maketitle 

\begin{abstract} 
The extraordinary magnetic Ap star HD\,75049 has been studied with data obtained 
with the ESO VLT and 2.2-m telescopes. Direct measurements reveal that the 
magnetic field modulus at maximum reaches 30\,kG. The star shows photometric, 
spectral and magnetic variability with a rotation period of 4.049\,d. Variations 
of the mean longitudinal magnetic field can be described to first order by a 
centred dipole model with an inclination $i = 25^\circ$, an obliquity $\beta = 
60^\circ$ and a polar field $B_p = 42$\,kG.
 The combination of the 
longitudinal and surface magnetic field measurements imply a radius of $R = 
1.7$\,R$_{\odot}$, suggesting the star is close to the zero-age main 
sequence. 
 HD\,75049 displays moderate 
overabundances of Si, Ti, Cr, Fe and large overabundances of rare earth elements. 
This star has the second strongest magnetic field of any main sequence star after 
Babcock's star, HD\,215441, which it rivals. 

\end{abstract} 


\section{Introduction} 

The complex processes of atomic diffusion -- radiative levitation and 
gravitational settling -- occur in upper main sequence stars, and are particularly 
evident in chemically peculiar A, B and F-type stars, collectively known as Ap 
stars. Accordingly, these stars are good targets to place observational 
constraints on atomic diffusion, which is crucial for deepening our understanding 
of these processes within the framework of studies of the structure and evolution 
of stars, and for more global analysis of galactic chemical evolution. The Ap 
stars rotate considerably more slowly than spectroscopically normal main sequence 
stars with similar effective temperatures. A fraction of Ap stars have extremely 
long rotation periods from tens of days to tens of years, and even of the order of 
a century in the case of $\gamma$\,Equ. There is no clear picture of the braking 
mechanism leading to such slow rotation, but the observational evidence strongly 
suggests that it is related to magnetic field (\citealt{Mathys97}; 
\citealt{Abt09}). 

Many Ap stars possess strong, predominantly dipolar magnetic fields. The first 
magnetic star, 78\,Vir, was found by \citet{babcock47}. Eleven years later 
\citet{babcock58} published the first catalogue of magnetic stars, and two years 
following that he discovered a huge 34\,kG  magnetic field in the star 
HD\,215441 \citep{babcock60}, known since then as ``Babcock's star''. Despite 
success in the years that followed in increasing the number known magnetic stars 
-- especially in last several years (for example \citealt{hubrigetal06}; 
\citealt{kudretal06}) -- HD\,215441 remains the record-holder for the main 
sequence star with the strongest magnetic field. The second strongest magnetic 
star, HD\,154708, was discovered by \citet{hubrigetal05}. Around the same time 
\citet{Kochukhov06} found HD\,137509 to have magnetic field modulus of 29\,kG. His 
result was in agreement with \citet{Mathys95} who discovered a very strong 
quadratic magnetic field up to 37\,kG in this star and mentioned that it has the 
one of the strongest magnetic fields. Surprisingly, the longitudinal magnetic 
field in HD\,137509 is not very large and varies just from $-1.25$ to $+2.35$\,kG 
\citep{Mathys91}. HD\,154708 still has the strongest magnetic field known among 
the cooler rapidly oscillating (roAp) stars \citep{kurtzetal06}. 

The origin of the magnetic fields in Ap stars and details of their peculiar 
properties are still mysteries, despite large efforts in the study of these stars. 
As usual in physics, analysis of extreme cases may provide key information, hence 
there is great interest in the stars with the strongest magnetic fields to help 
solve many problems in Ap stars. Recently, \citet{freyhammer08} discovered an 
extremely strong magnetic field in the Ap star HD\,75049. Many lines in the high 
resolution spectra are split into clear Zeeman patterns. From the value of 
splitting \citet{freyhammer08} estimated the photospheric magnetic field modulus 
to be about 30\,kG. These observations also revealed a reasonably short rotation 
period, and large variability both of the magnetic field and the spectral line 
intensities. Because of these interesting characteristics of this rare object, we 
have obtained a series of high resolution spectra and carried out circular 
spectropolarimetric observations with the VLT. We present in this paper the 
results of our analysis. 

\section{Observations and data reduction} 

To characterise the magnetic field of an Ap star, it is necessary to have magnetic 
measurements that cover the rotational phases of the star well. While we had a 
preliminary estimate for the rotation period of 4.05\,d determined by 
\citet{freyhammer08} from All Sky Automatic Survey (ASAS) photometric survey data 
\citep{Pojmanski02}, we were uncertain about this, so decided to spread new 
observations over half a year to be sure to determine the correct rotational 
period. {There is also a possible 5.28-yr period in the ASAS data that was 
noted by \citet{freyhammer08}. The nature of this long period is not clear, but it 
is certainly not a rotational period; the $v \sin i = 8.5$\,km\,s$^{-1}$  of 
HD\,75049 does not allow that. Further understanding of this period will require 
long-term observations.} 

A series high resolution spectra were obtained in service mode using UVES 
(Ultraviolet and Visual Echelle Spectrograph) on the ESO VLT (Very Large 
Telescope). Twelve spectra were collected between 2007 October 10 and 2008 April 
1. Two previous observations of this star with FEROS and UVES obtained in 2007 
February and March were also used for this analysis. These observations were made 
with echelle spectrographs over long spectral regions. For FEROS the spectral 
range is $3500 - 9220$\,\AA\ with a resolution $R = \lambda/\Delta\lambda = 
48\,000$; for UVES the range is $4970 - 7010$\,\AA\ with a resolution of $R = 
110\,000$. A journal of these observations is presented in Table\,1. Software 
packages provided by ESO (UVES and FEROS pipelines in the MIDAS environment) were 
used for the reduction and extraction of 1D spectra. 

\begin{table} 
\centering 
\caption{A journal of high resolution spectroscopic observations with UVES at the 
VLT and FEROS at the ESO 2.2-m telescope. The columns give the Barycentric Julian 
Date (BJD) of the middle of each exposure, the exposure time, the signal-to-noise 
ratio that was measured in sections of the continuum free of spectral lines. } 
\label{tab:hd75049_sp1}   
\begin{tabular}{lccc}   
\hline   
\hline   
\multicolumn{1}{c}{BJD}  &  exposure time   & S/N  & Instrument \\   
&  sec    & ratio  &  \\   
\hline   
2454141.66963   &   1500 &    190   &  FEROS  \\   
2454171.50422   & \phantom{0}900  &    180   &  UVES \\   
2454387.81593   &   1800 &    250   &  UVES \\   
2454432.82489   &   1800 &    240   &  UVES \\   
2454451.83366   &   1800 &    250   &  UVES  \\   
2454455.70226   &   1800 &    260   &  UVES  \\   
2454468.84589   &   1800 &    230   &  UVES  \\   
2454481.81465   &   1800 &    230   &  UVES  \\   
2454513.78892   &   1800 & \phantom{0}80   &  UVES  \\   
2454516.68904   &   1800 &    240   &  UVES  \\   
2454524.66461   &   1800 &    240   &  UVES  \\   
2454544.51608   &   1800 &    260   &  UVES  \\   
2454547.51127   &   1800 &    300   &  UVES  \\   
2454557.59095   &   1800 &    240   &  UVES  \\   
\hline   
\end{tabular}  
\end{table}  

We also obtained spectropolarimetric observations of HD\,75049 with the {\bf 
FO}cal {\bf R}educer and low dispersion {\bf S}pectrograph (FORS\,1) in the VLT in 
service mode during the time span 2007 December 30 to 2008 April 1. Observations 
with FORS\,1 spectra in circular polarisation are used for measurement of the 
longitudinal magnetic field, which corresponds to the line-of-sight component of 
the magnetic field vector integrated over the visible stellar hemisphere. This 
magnetic field parameter is sensitive to the magnetic field geometry and is very 
efficient for determining the rotational period. 

Table\,2 gives the Barycentric Julian Date of the middle of each observation, the 
exposure time, phase of the rotational period and the results of the longitudinal 
magnetic field measurements for three different lists of spectral lines or 
spectral regions. A detailed description of similar spectropolarimetric 
observations and its reductions is given in the recent paper by 
\citet{hubrigetal09}. 


\begin{table*} 
\centering 
\caption{Observations of HD\,75049 with FORS\,1. The columns give the Barycentric 
Julian Date (BJD) of the middle of each exposure, the rotational phase according 
to the ephemeris in eq.\,2, and the longitudinal magnetic field for two wavelength 
ranges of the spectrum and for the hydrogen lines with the error of measurement.} 
\label{tab:hd75049_lf}   
\begin{tabular}{ccrrr}   
\hline   
BJD  &    rotation & \multicolumn{3}{c}{longitudinal magnetic field $B_z$ (G)}   
\\   
& phase   &  \multicolumn{1}{c}{$3212-6215$\,\AA}     &    
\multicolumn{1}{c}{$3705-6215$ \,\AA}  &  \multicolumn{1}{c}{hydrogen lines}    \\   
\hline   
2454482.76787     &  0.3978  & $-8894 \pm   68$   &     $-8934 \pm   68$   &   
$-9744  \pm 132$       \\   
2454483.60947     &  0.6057  & $-9071 \pm   39$   &     $-9096 \pm   39$   &   
$-9938  \pm \phantom{0}74$        \\   
2454493.70464     &  0.0989  & $-905  \pm   32$   &     $-902  \pm   32$   &   
$-1475  \pm \phantom{0}61$        \\   
2454516.73363     &  0.7865  & $-4471 \pm   35$   &     $-4485 \pm   35$   &   
$-5064  \pm \phantom{0}65$        \\   
2454526.56586     &  0.2148  & $-3538 \pm   40$   &     $-3564 \pm   40$   &   
$-3970  \pm \phantom{0}74$        \\   
2454527.58159     &  0.4657  & $-9832 \pm   36$   &     $-9857 \pm   36$   &   
$-10529 \pm \phantom{0}72$        \\   
2454532.72343     &  0.7356  & $-5811 \pm   49$   &     $-5824 \pm   49$   &   
$-6367  \pm \phantom{0}91$        \\   
2454539.77171     &  0.4764  & $-9639 \pm   46$   &     $-9661 \pm   46$   &   
$-10334 \pm \phantom{0}94$        \\   
2454543.57545     &  0.4158  & $-9237 \pm   39$    &    $-9257 \pm   39$   &  
$-9941  \pm \phantom{0}75$        \\   
2454546.52195     &  0.1435  & $-1821 \pm   44$    &    $-1818 \pm   44$   &  
$-2364  \pm \phantom{0}86$        \\   
2454555.53106     &  0.3685  & $-8139 \pm   33$    &    $-8168 \pm   33$   &  
$-8528  \pm \phantom{0}64$        \\   
2454557.61944     &  0.8843  & $-1785 \pm   34$    &    $-1793 \pm   34$   &  
$-2397  \pm \phantom{0}62$        \\   
2454464.85968     &  0.9749  & $-640  \pm   36$    &    $-639  \pm   36$   &  
$-1415  \pm \phantom{0}68$        \\   
\hline   
\end{tabular}  
\end{table*}  

\begin{figure} 
\centering \epsfxsize 8cm\epsfbox{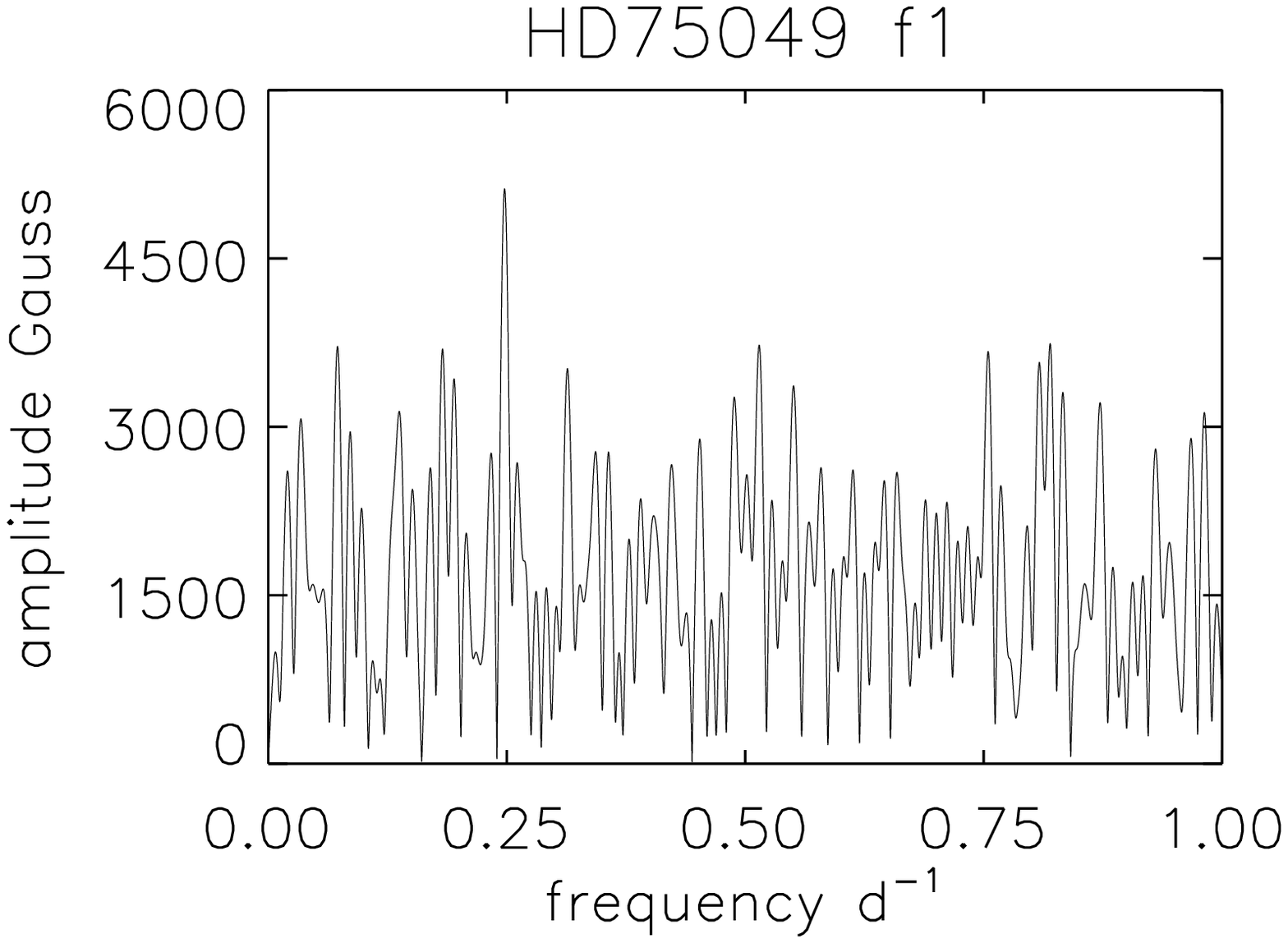} \epsfxsize 
8cm\epsfbox{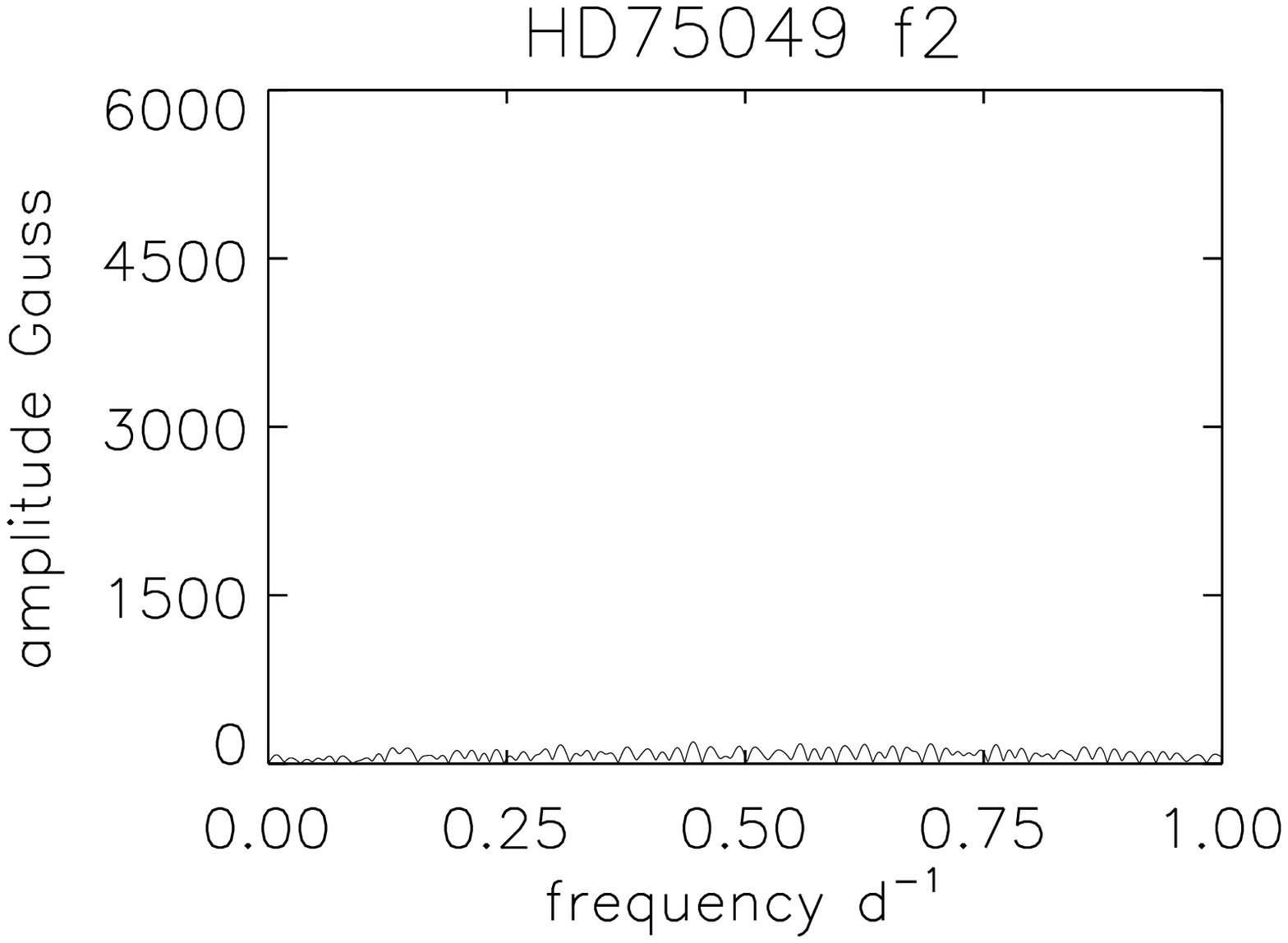} \epsfxsize 8cm\epsfbox{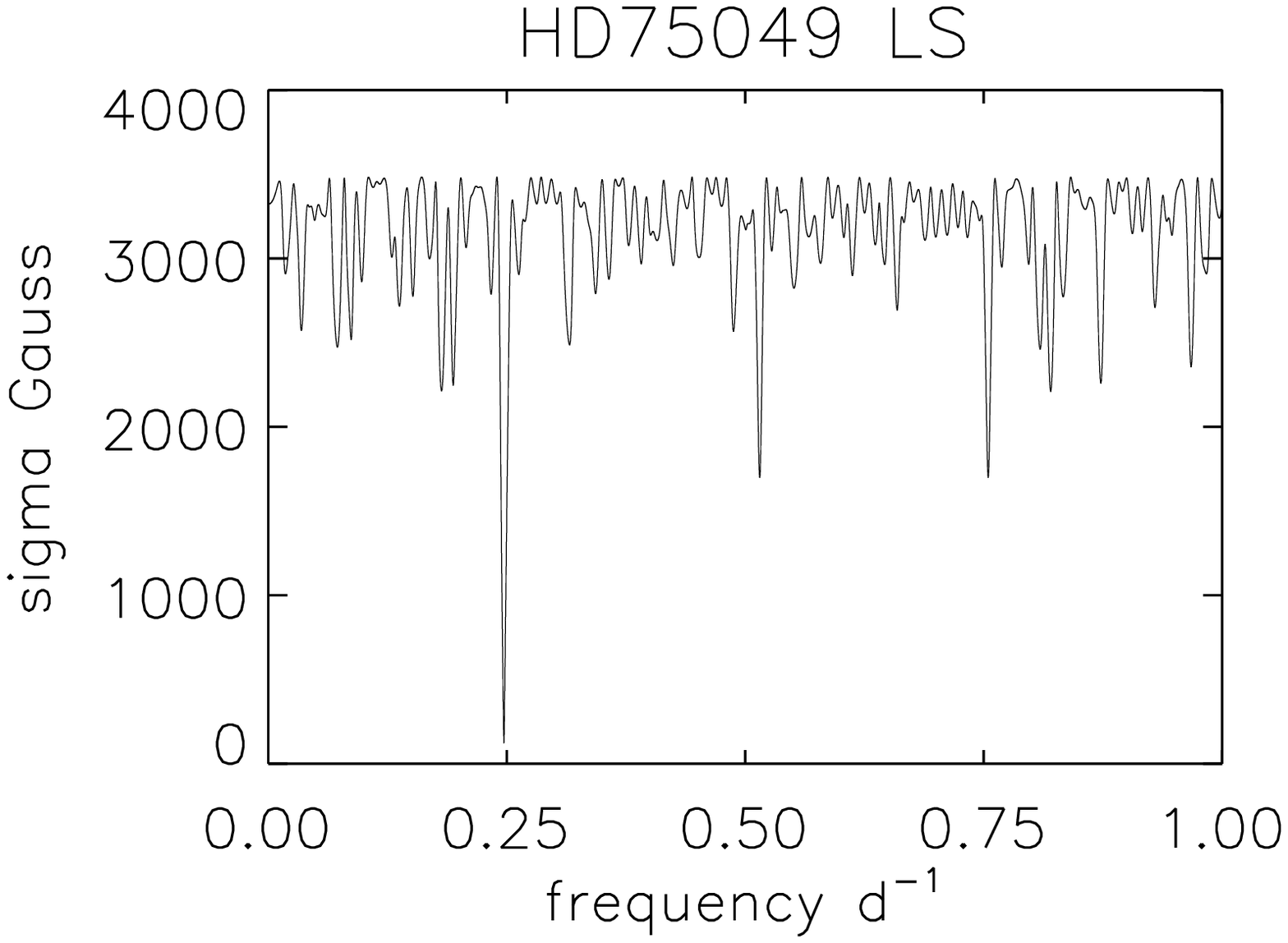} 
\caption{\label{fig:freq} Frequency analysis of 13 FORS1 measurements of $B_z$ 
with times in BJD. The top panel shows the amplitude spectrum with the highest 
peak at $f = 0.2477$\,d$^{- 1}$. The middle panel shows the residuals on the same 
scale after the highest peak has been refined by linear and nonlinear least 
squares fitting, and prewhitened from the data. It is clear that all of the 
apparent ``noise'' in the top panel is spectral window pattern. The bottom panel 
shows a least-squares fit of a sinusoid to the same data where the best-fitting 
frequency, $f = 0.2470$\,d$^{-1}$, is clear.} 
\end{figure} 

\section{Rotation period} 

The rotation period is an important parameter for the analysis of magnetic stars, 
which normally show photometric, spectral and magnetic variability with this 
period. In HD\,75049 we found a clear picture of variability in all three cases. 
As mentioned above, photometric variability was detected in ASAS data, and 
spectral and magnetic variability were found from high resolution spectra 
\citep{freyhammer08}. Our new longitudinal field measurements have allowed us to 
determine the rotation period with confidence and higher accuracy. 

Using a Discrete Fourier Transform \citep{Kurtz85} and the programme 
\textsc{period04} \citep{lenz05}, we found a highest peak in the amplitude 
spectrum of the FORS\,1 $B_z$ data (comprising 13 measurements) of $f = 
0.2477$\,d$^{-1}$; using least squares fitting of a sinusoid we found $f = 
0.2470$\,d$^{-1}$, as is seen in Fig.\,\ref{fig:freq}. We refined the fitted 
frequency and calculated uncertainties using nonlinear least-squares fitting, 
fitting the function 
\begin{equation} 
B_z = A_0 +A \cos[2\pi f(t - t_0) + \phi] 
\end{equation} 
which gives our final values: 

\vspace{2mm} 
\noindent $f = 0.246975 \pm 0.000005$\,d$^{-1}$\\ 
$P_{\rm rot} = 4.04899 \pm 0.00008$\,d\\ 
$A_0$ = $-5127 \pm 34$\,G\\ 
$A = 4824 \pm 40$\,G\\ 
$\phi = 0.000 \pm 0.010$\,rad, where\\ 
$t_0 = {\rm BJD}\,245\,4509.550 \pm 0.006$\\ 
$\sigma = 121$\,G 

\vspace{2mm} 
\noindent The standard deviation $\sigma$ is per measurement with respect to the 
fit. The error on $A_0$ is thus $\frac{\sigma}{\sqrt{13}}$, where $N = 13$ data 
points. The error on $t_0$ is the error in phase divided by $2 \pi f$. We thus get 
an ephemeris of 
\begin{equation} 
B_z ^{\rm max} = {\rm BJD}\,245\,4509.550 \pm 0.006 + 4.04899 \pm 0.00008 E. 
\end{equation} 
The value of the rotation period is consistent with that determined by 
\citep{freyhammer08} from ASAS photometry. 

\section{The stellar parameters} 

As an initial step to determine the fundamental stellar parameters we employed 
previous photometric observations. We used the $uvby\beta$ photometry by 
\citet{mart93} and the \textsc{UVBYBETA} program written by T.T. Moon and modified 
by \citet{napi93} based on the grid published in \citet{Moon85}. We derived an 
effective temperature, $T_{\rm eff} = 9600$\,K, and a gravity of $\log g = 4.47$ 
(cgs). { The surface gravity is an issue}, since it is known that 
photometric calibrations of the $c_1$ index, which are useful for determination of 
luminosity (hence surface gravity) for normal stars, are not entirely suitable for 
Ap stars, where they tend to underestimate luminosity because of heavy line 
blanketing in the $v$ filter. The effective temperature estimate, on the other 
hand, usually needs only a small correction (e.g. \citealt{Hubrig00}). The 
reliability of the gravity estimate from Str\"omgren photometry is discussed by 
\citet{North89}. {From Geneva 
photometry\footnote{http://obswww.unige.ch/gcpd/gcpd.html}   \citep{Mermilliod97}  
and using different calibrations  (\citealt{Cramer79}; \citealt{North90}; 
\citealt{Hauck93})  we obtained effective temperature estimates from 9300\,K to 
9900\,K. These, together with the estimate from $uvby\beta$, give an average 
photometrically-determined  value of $T_{\rm eff} = 9600 \pm 300\,$K.  
Interstellar reddening is negligible for this star. There is no clear 
evidence of any interstellar component in the Na\,\textsc{i} D1 and D2 lines, 
which show doublet splitting and belong to the star; any interstellar component is 
below 0.05\,\AA\ in equivalent width. This yields an upper limit to the colour 
excess of only $E_{B-V} < 0.02$ \citep{Munari97}.}

The Balmer lines profiles are sensitive to effective temperature and gravity. 
Reliable determination of the right continuum for them -- especially for echelle 
spectra -- is not an easy task and may be a source of scatter and errors. In the 
case of HD\,75049 magnetic broadening is large and needs to be taken into account 
in the Balmer lines, too. That small variations of Balmer lines may occur has long 
been known, as discussed recently by, e.g., \citet{Valyavin07} and 
\citet{Elkin08}. Balmer lines profiles of H$\alpha$, H$\beta$ and H$\gamma$ in the 
FEROS spectrum, H$\beta$ and H$\gamma$ in the FORS\,1 spectra, and H$\alpha$ in 
the UVES spectra were compared with synthetic profiles for best fits as a function 
of $T_{\rm eff}$ and $\log g$. An average gives $T_{\rm eff} = 9700 \pm 170\,$K 
and $\log g = 4.07 \pm 0.28$. The comparison between the observed and synthetic 
profiles of the Balmer lines gives acceptable agreement for a model with $T_{\rm 
eff} = 9600$\,K and $\log g = 4.0$. Synthetic calculations of the Balmer profiles 
were done with the \textsc{SYNTH} code of \citet{piskunov92}. Model atmospheres by 
\citet{kurucz79}, and from the NEMO database \citep{heiteretal02} were used.

\section{Magnetic field} 

\subsection{Longitudinal field}
From observations with FORS\,1 the variable mean longitudinal magnetic field 
$B_{z}$ was determined. Table\,2 gives the results obtained for three methods of 
calculation when various lists of spectral lines and spectral regions are used. 
There are no differences between $B_{z}$ obtained from all spectral lines in two 
spectral regions, $3212 - 6215$\,\AA\ and the shorter $3705 - 6215$\,\AA. But 
there are shifts of around 600\,G between the latter and measurements from the 
hydrogen lines. These differences most likely reflect the uncertainties involved 
in the interpretation of the observed circular polarization signal in terms of a 
magnetic field, which arise from the statistical nature of the measurement (the 
considered spectral ranges include many lines of different elements, which have 
different magnetic sensitivities) or from the complexity of the physical 
foundations of the treatment of the formation of hydrogen lines in A-type star 
atmospheres in the presence of a magnetic field \citep{Mathys00}. Differences in 
the non-uniform distribution of chemical elements in the stellar atmosphere may 
also contribute, but seem unlikely to represent the dominant effect. The phase 
curves for all three methods are similar and show sine curves. The distribution of 
hydrogen in the stellar atmosphere is more homogeneous than for other chemical 
species. Therefore we used results from hydrogen ({last column in 
Table\,2})  for further analysis. 

\subsection{Mean magnetic field modulus}

\begin{table*} 
\centering 
\caption{\label{tab:hd75049_B} The magnetic field modulus $\langle B \rangle$ (kG) 
in HD\,75049 determined from resolved Zeeman components in some selected spectral 
lines. Because of the non-uniform surface abundance distributions of various ions, 
some of the differences in field strength at specific rotation phases may be 
attributable to the distribution of elements. Less reliable results are identified 
by question marks. } 
\begin{tabular}{cccccccc}   
\hline   
BJD &  phase & Fe\,\textsc{ii}  &   Nd\,\textsc{iii} & Cr\,\textsc{ii} &   
Nd\,\textsc{iii}    & Nd \,\textsc{iii}   &     
Eu\,\textsc{ii} \\   
&       & 5018\,\AA & 5050\,\AA  &  5237\,\AA  &  6145\,\AA   &  6327\,\AA    &     
6437\,\AA       \\   
\hline   
2454141.66963 &  0.1550 &  26.08   &  26.58  &  26.83   &  26.67      &  26.34         
&   29.49   \\   
2454171.50422 &  0.5234 &  29.54   &  28.76  &  28.23   &  29.68      &  28.44         
&   30.10   \\   
2454387.81593 &  0.9470 &  25.66   &  26.58  &  25.91   &  28.32      &  26.05         
&   28.78   \\   
2454432.82489 &  0.0631 &  24.65   &  25.19? &  25.14   &  25.72      &  26.08         
&   29.05   \\   
2454451.83366 &  0.7578 &  27.99   &  27.98  &  27.68   &  29.12      &  28.01         
&   29.62   \\   
2454455.70226 &  0.7133 &  28.46   &  28.29  &  27.91   &  29.12      &  28.38         
&   29.87   \\   
2454468.84589 &  0.9594 &  25.24   &  25.82  &  25.64   &  28.10      &  26.08         
&   28.67   \\   
2454481.81465 &  0.1624 &  26.83   &  26.64  &  27.97   &  27.13      &  28.73         
&   29.64   \\   
2454513.78892 &  0.0593 &  24.25   &  24.74\rlap{?} &  24.84\rlap{?}  &  
23.27\rlap{?}     &                
&           \\   
2454516.68904 &  0.7755 &  27.73   &  27.69  &  27.48   &  29.20      &  27.60         
&   29.52   \\   
2454524.66461 &  0.7453 &  28.04   &  27.98  &  27.83   &  29.26      &  28.21         
&   29.62   \\   
2454544.51608 &  0.6481 &  28.96   &  28.60  &  28.18   &  29.40      &  28.38         
&   30.06   \\   
2454547.51127 &  0.3878 &  29.51   &  28.44  &  28.57   &  29.63      &  29.10         
&   30.04   \\   
2454557.59095 &  0.8773 &  26.45   &  26.49  &  26.71   &  28.77      &  26.77         
&   29.98   \\   
\hline   
\end{tabular}  
\end{table*}  

\begin{figure} 
\begin{center} 
\hfil \epsfxsize 8.0cm\epsfbox{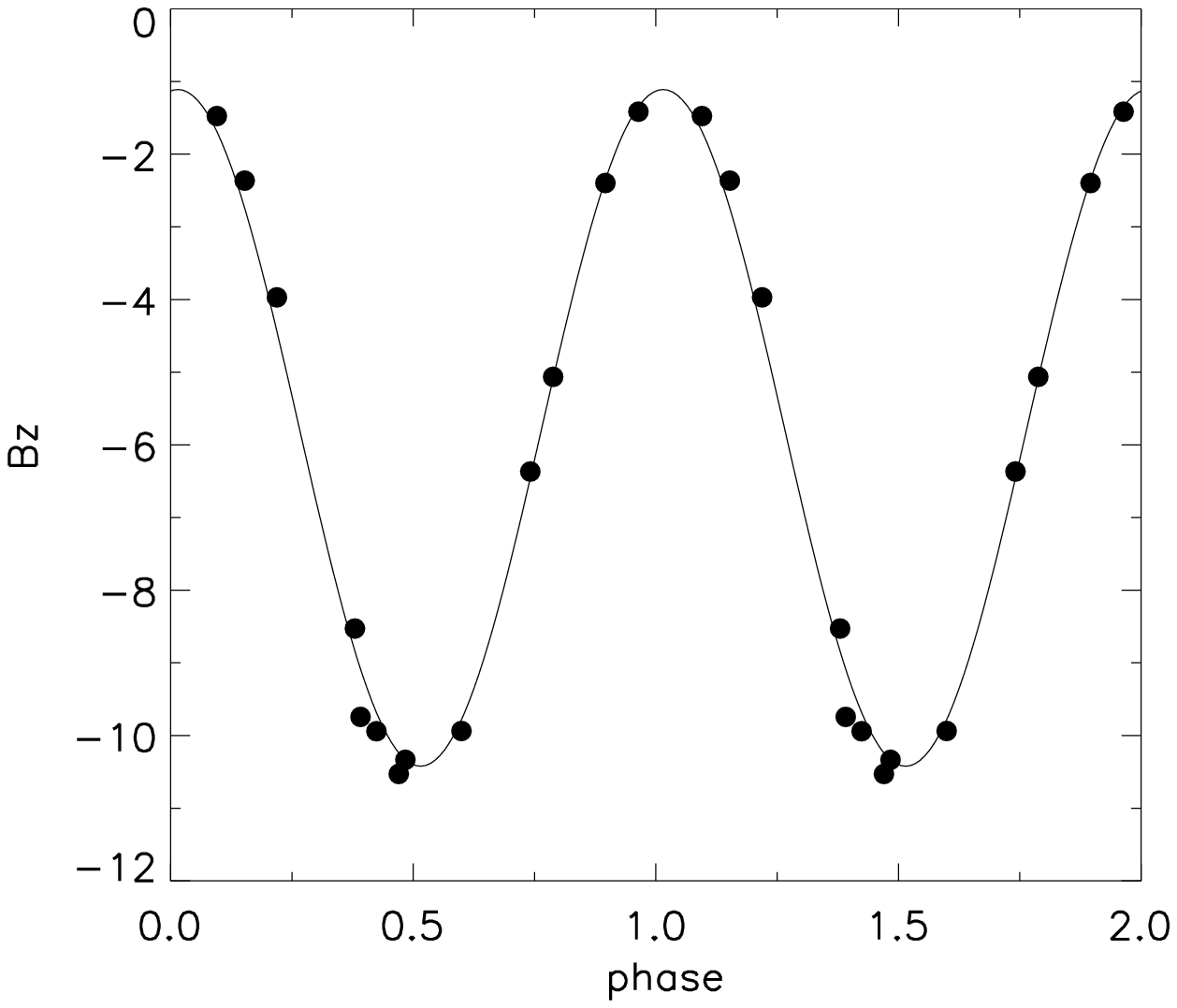} 
\hfil \epsfxsize 8.0cm\epsfbox{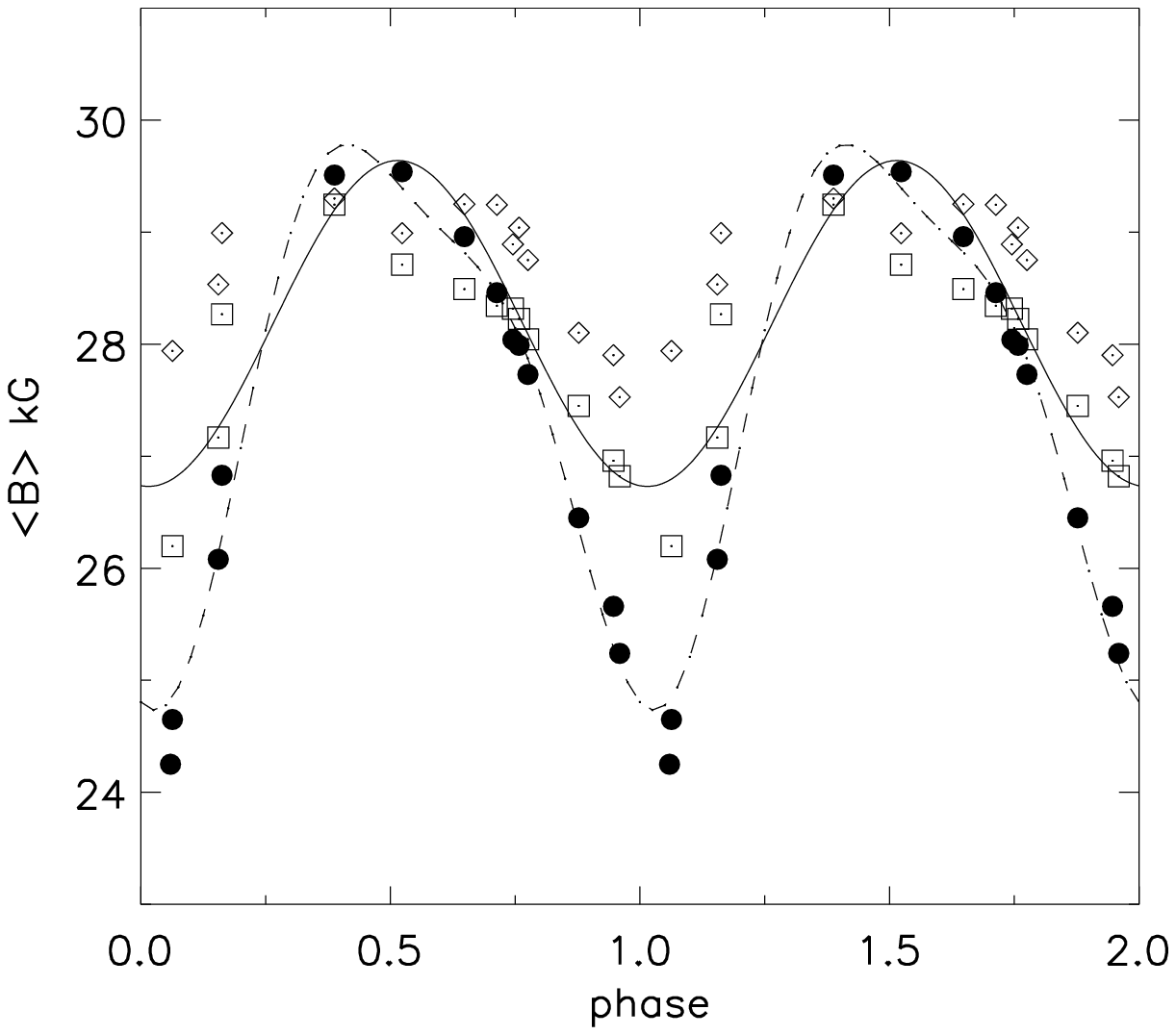} 
\caption{\label{fig:bzbm} Top panel: The variation of the longitudinal magnetic 
field $B_{z}$ with rotational period for HD\,75049. Filled circles are 
observations with FORS\,1 using the hydrogen lines. The curve is a centred dipole 
model for $B_p = 42$\,kG, $i=25^\circ$ and $\beta = 60^\circ$. Bottom panel: The 
variation of the magnetic field modulus $\langle B \rangle$ with rotational period 
for HD\,75049 from our UVES and FEROS data. Different symbols are used to 
represent measurements performed using lines of different ions: filled circles for 
Fe\,\textsc{ii} $\lambda\lambda$\,5018\,\AA, open squares for Nd\,\textsc{iii} 
using four lines at 
$\lambda\lambda$\,5050\,\AA, 5845\,\AA, 6145\,\AA\ and 6327\,\AA, and diamonds for 
Eu\,\textsc{ii} using two lines at $\lambda\lambda$\,6049\,\AA\ and 6437\,\AA. The 
solid line is the same model as for upper panel. The dashed curve is a 
least-squares fit of a sinusoid plus first harmonic to the Fe\,\textsc{ii} data.} 
\end{center}
\end{figure} 

The mean magnetic field modulus $\langle B \rangle$ was determined from high 
resolution spectra using resolved Zeeman components of several spectral lines. 
Using Gaussian fitting we determined the centres of shifted Zeeman $\sigma$ 
components. The distance between these two components in a spectral line is 
proportional to value of magnetic field modulus $\langle B \rangle$ 
(\citealt{Mathys97}): 

\begin{equation} 
\Delta\lambda = 9.34 \times 10^{-13} g_{\rm eff} 
\langle B \rangle \lambda_{0} ^{2} 
\label{eq:B} 
\end{equation} 

\noindent where wavelength is measured in \AA\ and $g_{\rm eff}$ is an effective 
Land\'e factor. This relation is valid in fairly general conditions for lines with 
doublet or triplet Zeeman patterns; in the general case of more complex, anomalous 
patterns, it represents only a first approximation. Furthermore, it is based on 
the assumption that the Doppler effect due to stellar rotation is small compared 
to the magnetic splitting of the line. The latter is not strictly fulfilled in 
HD\,75049. The resulting asymmetries of the line components, mostly seen between 
phases 0.85 and 0.35, further hamper the accurate determination of the field 
modulus. 

Since the star is spotted, the actual field strength measured is somewhat 
dependent on the surface distribution of ions from whose lines the measurements 
were made. Thus we determined the maximum field strength according to the set of 
spectral lines used. Table\,3 presents our measurements of $\langle B \rangle$ for 
some spectral lines. Because of the exceptional strength of the magnetic field, 
the spectral line of Fe\,\textsc{ii} 6149\,\AA, which lends itself well to the 
accurate determination of the mean field modulus in Ap stars and has been 
frequently used to this effect, is strongly distorted by the partial Paschen-Back 
effect, and its splitting into several components cannot be readily interpreted, 
so that we did not consider it. In such a strong field, the majority of spectral 
lines show resolved Zeeman splitting, but many are not strong enough for precise 
measurements. Many other lines are blends and cannot be used either. The most 
precise and reliable line splitting measurements were obtained for the strong line 
Fe~{\sc ii} $\lambda$\,5018\,\AA. Their mean uncertainties can be estimated from 
the standard deviation of the individual measurements about a least-squares fit of 
the superposition of a sinusoid and of its first harmonic against rotation phase. 
They are found in this way to be of the order of 0.27\,kG.

\subsection{Dipole model}

From \cite{Harmanec88} we estimate for a main sequence A star with $T_{\rm eff} = 
9600$\,K that the stellar radius will be around 2.1\,R$_{\odot}$. This value 
agrees well with other published estimations of stellar radii. {While  
main sequence stars with a similar temperature may range in radius between 
1.7\,R$_{\odot}$ and 
5\,R$_{\odot}$ from the zero-age main sequence to the terminal-age main sequence, 
a radius more than 2.4\,R$_{\odot}$ is ruled out for HD\,75049, as it requires an 
implausibly high polar magnetic field in comparison with the observed magnetic 
field modulus for the dipole model discussed in the next section. The 
range of possible radius for HD\,75049 is therefore between 1.7\,R$_{\odot}$ and 
2.4\,R$_{\odot}$.}  

The spectral lines with zero Land\'e factor, and resolved Zeeman components 
of lines with large Land\'e factors, give $v \sin i = 8.5 \pm 
1.0$\,km\,s$^{-1}$.
From the relation: 
$ v_{eq} = \frac{50.59R}{P} $, 
\noindent where $P$ is measured in days, we derive a rotational inclination of 
{ $i = 23.6^\circ \pm 3.0^\circ$  and an equatorial velocity $v_{\rm eq} = 
21.2$\,km\,s$^{-1}$ for 1.7\,R$_{\odot}$, and $i = 16.5^\circ \pm 2.0^\circ$ and 
$v_{\rm eq} = 30.0$\,km\,s$^{-1}$ for 2.4\,R$_{\odot}$.} 
 
From least squares fitting with the Period04 program \citep{lenz05}, for the 
hydrogen lines we have an 
average $B_{z} = -5.76$\,kG which varies with an amplitude 
of $4.89 \pm 0.06$\,kG. From expressions for an oblique rotator (e.g., 
\citealt{Preston67}) { we find a magnetic obliquity in the range $\beta = 
63^\circ \pm 3.0^\circ$  for star with radius 1.7\,R$_{\odot}$  to $\beta = 
71^\circ \pm 3.0^\circ$ for  2.4\,R$_{\odot}$}.{Following the approach of 
\citep{Stibbs50} and \citep{Preston70} we calculated a grid of models using the 
above parameter range and found a best fit for $B_{p} = 42 \pm 2$\,kG, $i = 
25^\circ \pm 3^\circ$ and $\beta = 60^\circ \pm 3^\circ$, values that are suitable 
for a stellar radius of 
1.7\,R$_{\odot}$. The variation of the longitudinal magnetic 
field $B_{z}$ with the rotational period and selected model fitting is shown in 
the top panel of Fig.\,\ref{fig:bzbm}.  We can get an equally good fit to the top 
panel of Fig.\,\ref{fig:bzbm} with the parameters, $R = 2.4$\,R$_{\odot}$, $i = 
16.5^\circ$, $\beta = 71^\circ$ and $B_{p} = 60$\,kG, but then the model curve for 
$\langle B \rangle$ in the bottom panel has a minimum of 38\,kG, which is clearly 
wrong. Under the assumption that the field geometry is dipolar, 
the implication of this is that HD\,75049 is close to the 
zero-age main sequence. }

{Variations of $\langle B \rangle$ show less symmetric curves than that 
for $B_z$. To fit the $\langle B \rangle$ curve for Fe\,\textsc{ii} 
5018\,\AA\ we had to employ a sine curve with its first harmonic. 
Our centred dipole model does not fit the variation of this Fe\,\textsc{ii} 
line but is acceptable for lines of Nd~{\sc iii} and Eu~{\sc ii}, considering 
the errors of measurement and nonuniform distribution of chemical elements. 
Otherwise the magnetic field geometry is more complicated than a simple dipole, 
as is typical for many magnetic Ap stars (e.g., \citealt{Landstreet00}) .} 

\begin{figure*} 
\resizebox{\hsize}{!}{\includegraphics{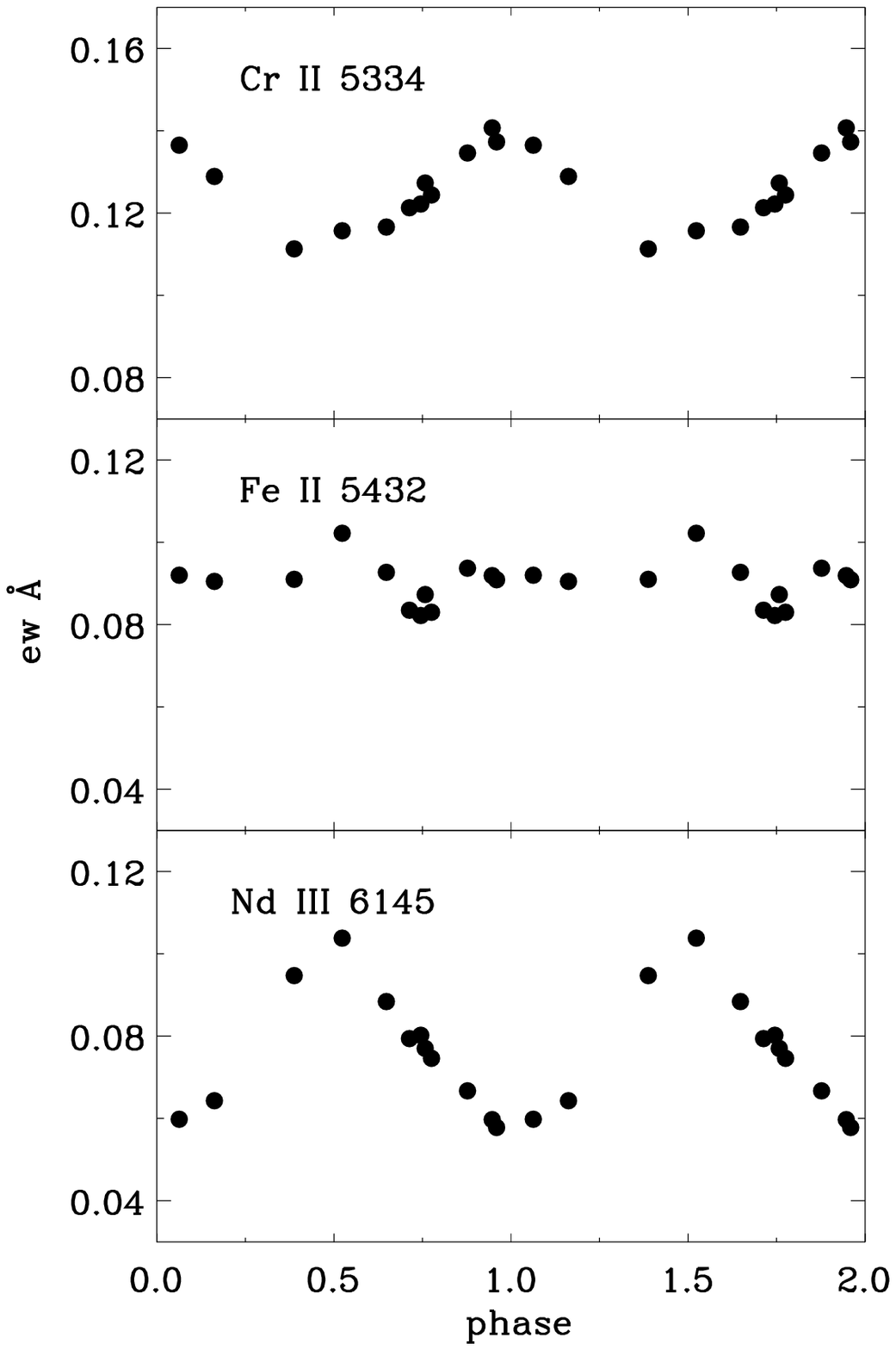}\includegraphics{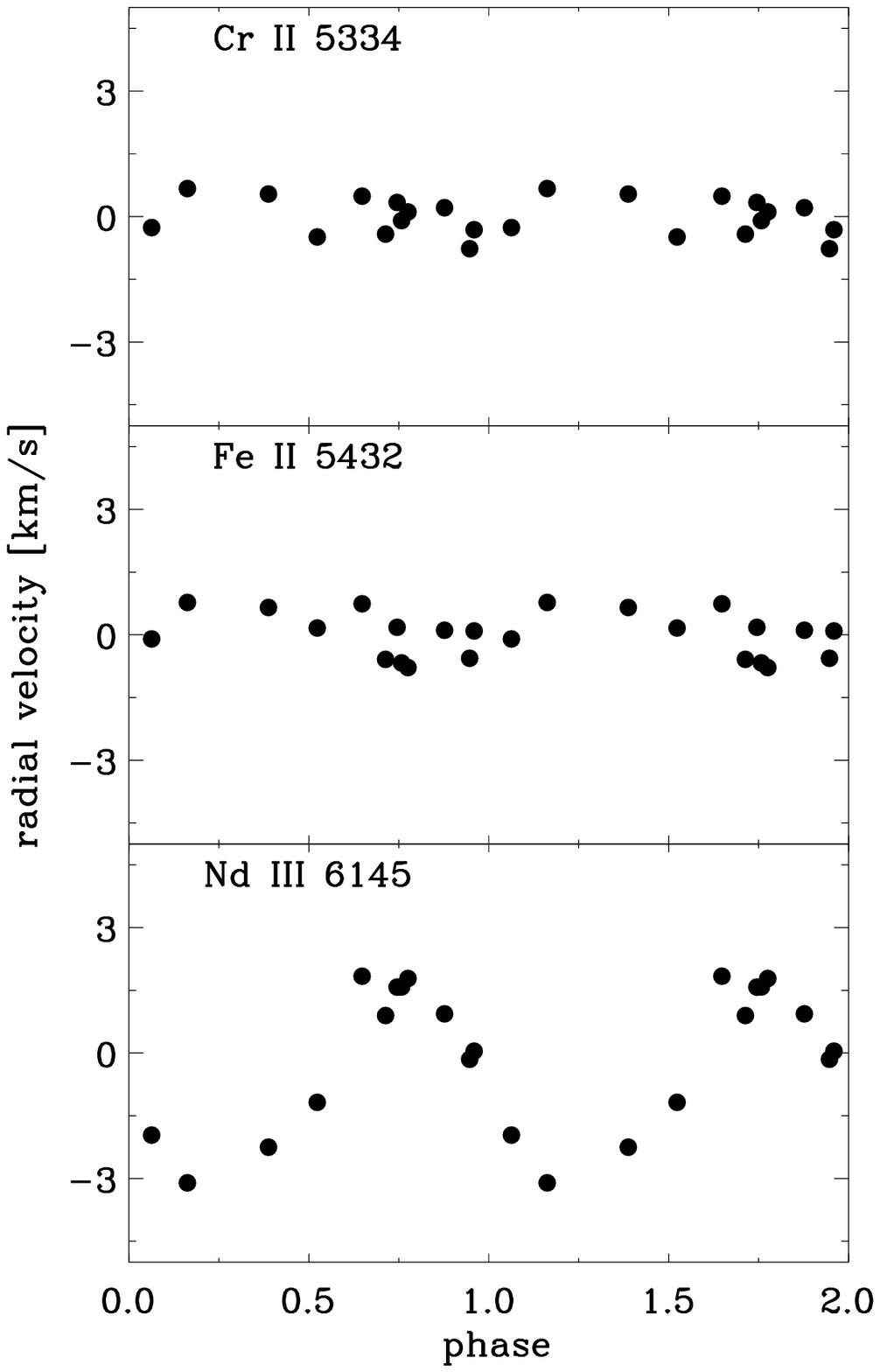}}
\caption{\label{fig:ew_rv_var} Variation with rotation phase of the equivalent 
width (left) and the radial velocity (right) of the lines Cr~{\sc ii} 
$\lambda$\,5334\,\AA, Fe~{\sc ii} $\lambda$\,5432\,\AA, and Nd~{\sc iii} 
$\lambda$\,6145\,\AA.} 
\end{figure*} 

\subsection{ Correlation between magnetic and spectral variability}

One sees in Fig.\,\ref{fig:bzbm} that the field modulus measurements obtained from 
the three sets of analysed lines show different amplitudes of variation. The 
Fe-based measurements vary with the largest amplitude, while the lowest variation 
amplitude is observed for $\langle B\rangle$ values derived from consideration of 
the Eu lines. These differences may find their origin in part in differences in 
the non-uniform distribution of the various elements of interest over the stellar 
surface. In order to test this interpretation, we have checked the dependence on 
rotation phase of the equivalent widths and radial velocities of lines of the 
considered elements. The examples in Fig.\,\ref{fig:ew_rv_var} illustrate the 
observed behaviours. The spectral lines that are shown, Cr~{\sc ii} 
$\lambda$\,5334\,\AA, Fe~{\sc ii} $\lambda$\,5432\,\AA\ and Nd~{\sc iii} 
$\lambda$\,6145\,\AA, are representative of all the reasonably unblended lines of 
the elements from which they arise. Equivalent width variations are definitely 
observed for Cr and Nd, but they are not definitely detected for Fe. The Nd lines 
also show radial velocity variations, while our data show no clear indications of 
such variations for Cr and Fe. Even after averaging the radial velocity 
measurements of several lines of these elements, no convincing evidence for 
variability was found. 

For Nd, maximum equivalent width occurs close to phase 0.5, hence to the negative 
extremum of the longitudinal field, and to the maximum of the field modulus. The 
radial velocities of the Nd lines vary in phase quadrature with their equivalent 
widths; the lines are blueshifted between phases 0 and 0.5, and redshifted between 
phases 0.5 and 1.0. This is consistent with the presence of a region of enhanced 
abundance (``spot'') of Nd around the negative magnetic pole. The same conclusion 
applies to Eu, whose lines show similar variability to those of Nd. 

The equivalent widths of the Cr lines show the opposite variation, reaching their 
minimum close to phase 0.5, and being largest close to phase 0. Any radial 
velocity variations that they may show do not clearly stand out of the measurement 
noise, so that their amplitude is definitely much lower (by a factor of 5 or more) 
than for Nd lines. Yet, this may not be inconsistent with equivalent width 
variations being due to a non-uniform distribution of Cr over the stellar surface 
such that the abundance of this element is minimum around the negative pole, and 
increases away from it. Possible configurations include a band of enhanced Cr 
abundance along the magnetic equator. With a distribution of this type, one can 
simultaneously observe on the visible stellar hemisphere approaching and receding 
Cr-rich regions, whose contributions to the radial velocity of Cr spectral lines 
mostly averages out in disk-integrated observations. 

Finally, Fe lines do not show any definite variability of their equivalent widths 
or radial velocities, indicating that any abundance inhomogeneity in the 
distribution of this element on the visible part of the star must be moderate at 
most. Si behaves like Fe. 

Thus we are left with the qualitative picture of a star showing a concentration of 
Nd and Eu and a (relative) depletion of Cr around its negative magnetic pole, with 
a comparatively uniform distribution of Fe and Si over its surface. One should 
keep in mind, though, that due to the low inclination of the rotation axis on the 
line of sight, a large fraction of the stellar surface is never observed. 

The non-uniform distribution of Nd over the stellar surface may account for part 
of the differences between the values of $\langle B\rangle$ that are derived from 
measurements of its lines, and from those of Fe, whose abundance appears much more 
constant across the star. However, it is unlikely to be the only factor, or even 
the main one. If it were, one would, for instance, expect greater values of the 
field modulus to be determined from Nd lines than from Fe lines close to phase 
0.5; the opposite is observed in Fig.\,\ref{fig:bzbm}. Most likely, the origin of 
the difference should be sought elsewhere. In particular, one should bear in mind 
that Nd lines are subject to hyperfine structure. This introduces severe 
complications in the treatment of their formation in the presence of a strong 
magnetic field, as illustrated, e.g., in \citet{landi75}. Accordingly, the 
physical meaning of the value of $\langle B\rangle$ that is obtained by 
application of Eq.~(\ref{eq:B}) to the observed splitting of Nd lines is not fully 
clear. In particular, one cannot a priori expect it to provide a measurement of 
the mean magnetic field modulus that is consistent with its determinations from 
analysis of Fe lines, which rests on a firmer physical basis. The same applies to 
the usage of Eu lines to measure $\langle B\rangle$. The differences seen in 
Fig.\,\ref{fig:bzbm} between the variation curves of this field moment resulting 
from measurements of the Nd~{\sc iii} and Eu~{\sc ii} lines would actually be very 
difficult to understand in terms of abundance inhomogeneities since the equivalent 
width and radial velocity behaviours indicate that, to first order, the 
distribution of both elements over the stellar surface is similar. On the other 
hand, around phase 0, the lines Eu~{\sc ii} $\lambda\lambda$\,6049\,\AA \ and 
6437\,\AA\ become very weak and shallow, so that the uncertainties of the $\langle 
B\rangle$ determinations based on them become very large, and the values obtained 
for this field moment between phases $\sim 0.9$ and $\sim 0.2$ should be 
considered with appropriate caution. 

\section{Asymmetry of spectral lines} 

\begin{figure*} 
\resizebox{\hsize}{!}{\includegraphics{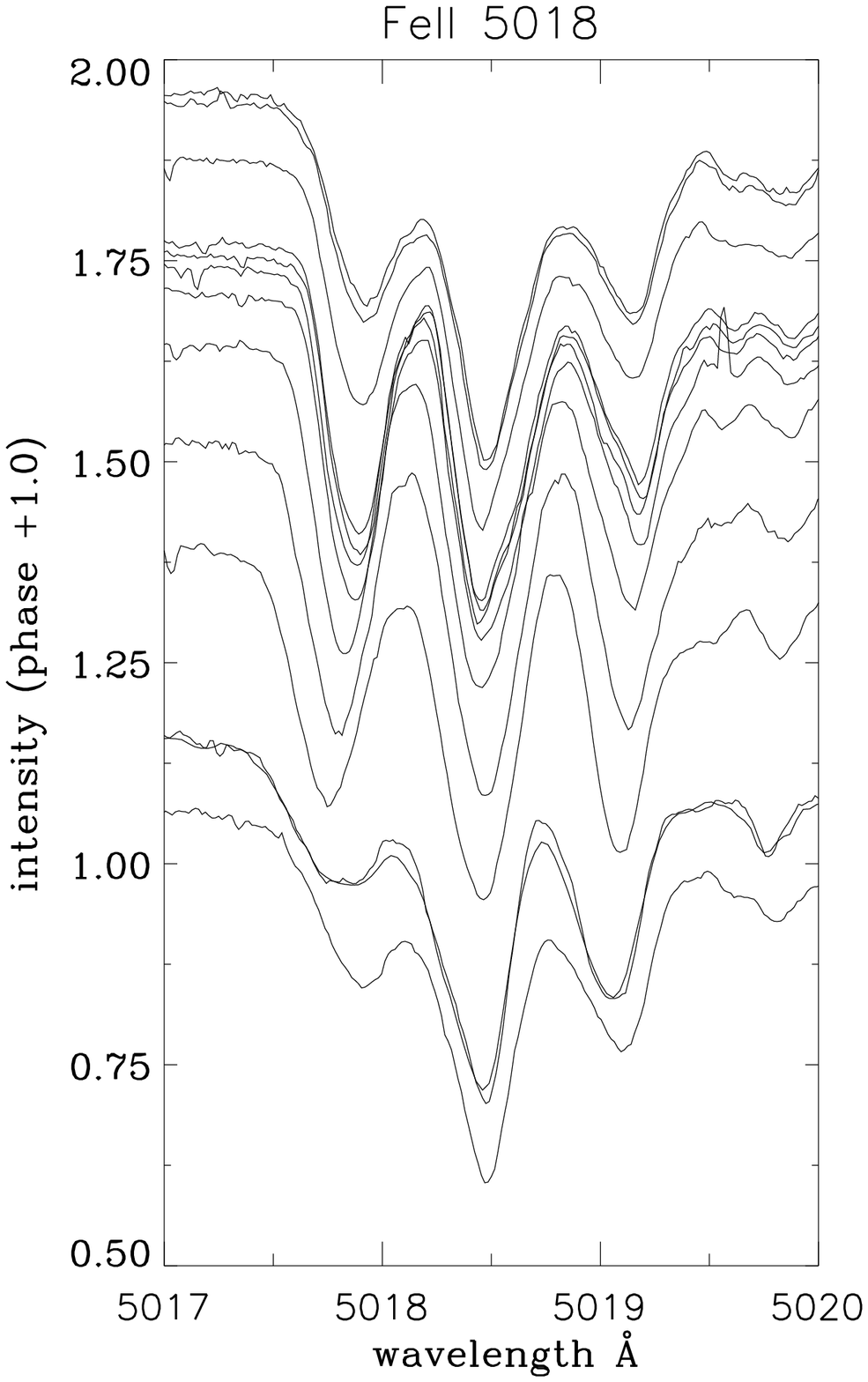}\includegraphics{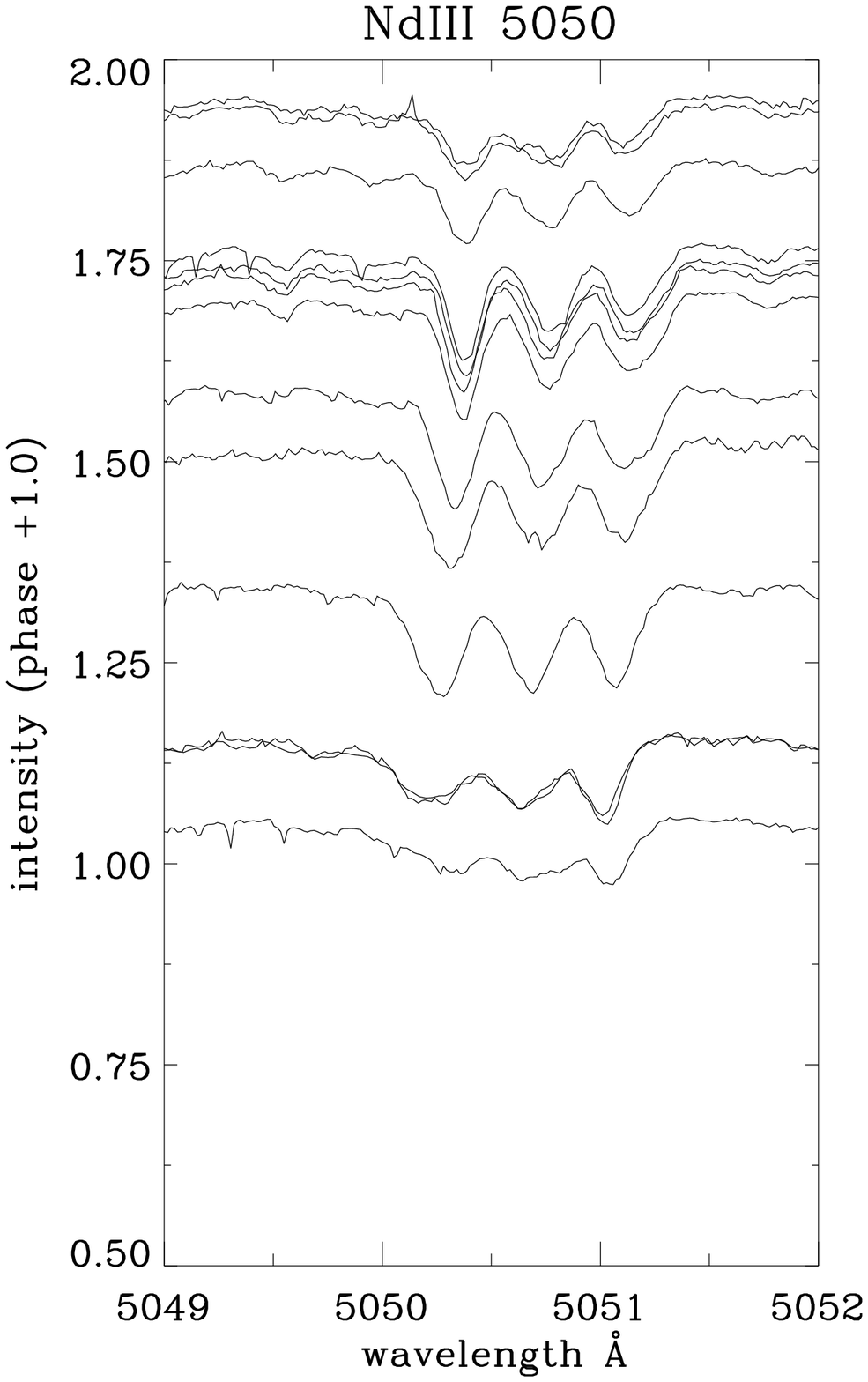}
\includegraphics{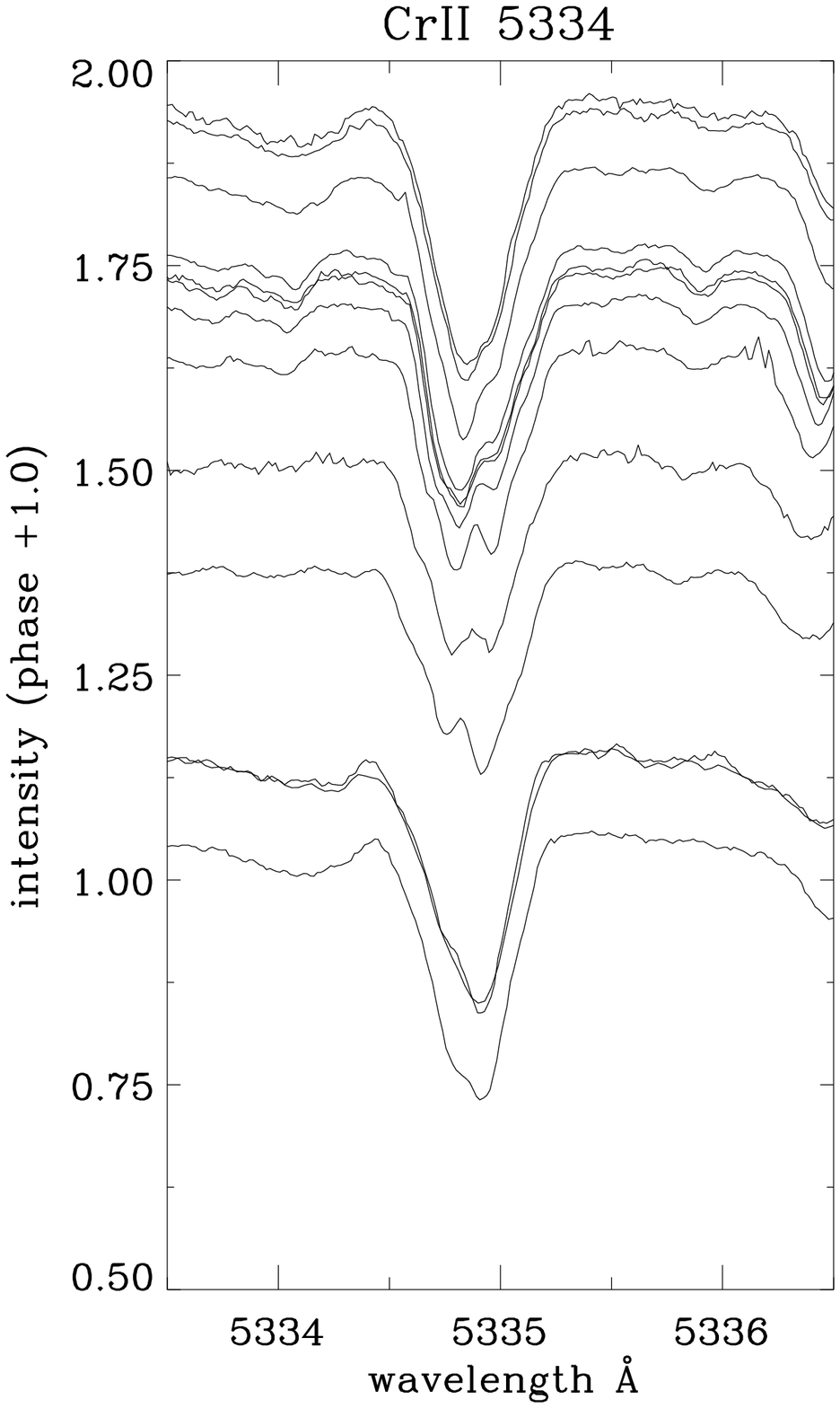}} 
\caption{\label{fig:asym} Variation with rotation phase of the observed profiles 
of the spectral lines Fe~{\sc ii} $\lambda$\,5018\,\AA\, Nd~{\sc iii} 
$\lambda$\,5050\,\AA, and Cr~{\sc ii} $\lambda$\,5334\,\AA. Each spectrum is 
shifted vertically from the normalised continuum to a value corresponding to (1 + 
rotation phase), hence the ordinate shows both continuum and rotation phase.} 
\end{figure*}

Spectral lines in HD\,75049 show significant variability with rotation period, and 
for many lines we also note asymmetry in the Zeeman patterns, as can be seen in 
Fig.\,\ref{fig:asym}. Shifted $\sigma$ components normally are symmetric with 
respect to the line centre, according to classical Zeeman theory. Observations of 
magnetic Ap stars reveal more complicated behaviours of the Zeeman patterns. 
Spectral line asymmetry is often visible in high resolution spectra of Ap stars. 
While non-uniform distribution of chemical species on the stellar surface 
contributes to this, the main source of line asymmetry more often is the variable 
combination of Zeeman and Doppler effects across the stellar surface. Indeed, due 
to the large-scale structure of the magnetic field, the Zeeman and Doppler shifts 
are correlated across the stellar disk, and this correlation is reflected in the 
shapes of the disk-integrated spectral lines. 

The observed presence of this effect in HD\,75049 is particularly plausible, 
considering that this star has a very strong magnetic field, and a large $v\,\sin 
i$ for a star with magnetically resolved lines. That it is the dominant effect for 
explanation of the origin of the line asymmetries can be inferred from 
consideration of Fig.\,\ref{fig:asym}. Representative lines of each of the three 
elements Fe, Cr, and Nd are seen to show qualitatively similar asymmetries: a red 
$\sigma$ component (or red wing) shallower and broader than its blue counterpart 
at phases comprised between $\sim$0.6 and $\sim$0.9, and vice-versa in the phase 
range $\sim$$0.1 - 0.4$. Since the distributions of Fe, Cr and Nd over the star 
differ considerably from each other, the similarity of the line asymmetries for 
all of them cannot arise primarily from their inhomgeneities. Also, because the 
magnetic field of HD\,75049 is very strong, partial Paschen-Back effect may 
contribute to asymmetries of a significant number of lines \citep{Mathys90}, but 
it affects different transitions to a variable extent and in qualitatively 
different ways, so that it cannot account for the consistent behaviours 
illustrated in Fig.\,\ref{fig:asym}. 

\section{Chemical abundances}

\begin{table*} 
\caption[]{Chemical abundances for HD\,75049 for selected elements, and their 
corresponding solar abundances \citep{asplundetal05}. The errors quoted are 
internal standard deviations for the set of lines measured. Only an upper limit 
could be placed on the abundance of Nd\,\textsc{ii}. Columns 3 and 5 give the 
number of lines used, ${\rm  N}_{\rm lines}$, in each case. } 
\begin{center}   
\begin{tabular}{lccccc}  
\hline    
\multicolumn{1}{c}{Ion}  &  
\multicolumn{1}{c}{$\log\,N/N_{\rm tot}$}  & 
\multicolumn{1}{c}{${\rm  N}_{\rm lines}$}  &  
\multicolumn{1}{c}{$\log\,N/N_{\rm tot}$}  &  
\multicolumn{1}{c}{${\rm  N}_{\rm lines}$}  &  
\multicolumn{1}{c}{$\log\,N/N_{\rm tot}$} \\   
& \multicolumn{1}{c} {$\phi=0.388$}    &
& \multicolumn{1}{c}{$\phi=0.063$}    &
& \multicolumn{1}{c}{Sun} \\    
\hline    
Si\,\textsc{ii}  & $-3.70 \pm 0.05$ &  2 & $-3.55 \pm 0.05$ & 2 & $-4.49$ \\ 
Ti\,\textsc{ii}  & $-6.04 \pm 0.10$ &  2 & $-5.85 \pm 0.15$ & 2 & $-7.10$ \\ 
Cr\,\textsc{ii}  & $-4.73 \pm 0.15$ & 14 & $-4.53 \pm 0.13$ & 9 & $-6.36$ \\ 
Fe\,\textsc{i}   & $-3.93 \pm 0.17$ & 10 & $-4.10 \pm 0.29$ & 4 & $-4.55$ \\ 
Fe\,\textsc{ii}  & $-3.89 \pm 0.15$ & 17 & $-3.98 \pm 0.15$ & 6 & $-4.55$ \\ 
La\,\textsc{ii}  & $-7.15 \pm 0.21$ & 16 & $-7.23 \pm 0.18$ & 6 & $-10.87$ \\ 
Ce\,\textsc{ii}  & $-7.48 \pm 0.16$ &  5 & $-7.95 \pm 0.05$ & 4 & $-10.42$ \\ 
Pr\,\textsc{iii} & $-8.07 \pm 0.15$ &  4 & $-8.10 \pm 0.16$ & 4 & $-11.29$ \\ 
Nd\,\textsc{iii} & $-7.27 \pm 0.05$ &  3 & $-7.57 \pm 0.12$ & 3 & $-10.55$ \\ 
Nd\,\textsc{ii}  & $< -8.40 $       &  3 & $< -7.98$        & 3 & $-10.55$ \\ 
Eu\,\textsc{ii}  & $-7.71 \pm 0.19$ &  4 & $-8.10 \pm 0.29$ & 4 & $-11.48$ \\ 
\hline    
\end{tabular}    
\label{abund}  
\end{center} 
\end{table*}    

Large overabundances of some chemical elements, especially of rare earth elements, 
is one of the defining characteristics of Ap stars. Various chemical elements 
concentrate in different places on the surface of the star. There is a tendency 
for rare earth elements to concentrate around the magnetic poles in some sort of 
spots. The size and number of spots may vary (\citealt{Kochukhov04}; 
\citealt{Freyhammer09}). The situation is even more complex since the atmospheres 
of Ap stars also show vertical stratification (e.g., \citealt{Babel92}; 
\citealt{Wade01}; \citealt{koch09}) where various elements accumulate in different 
layers in the atmosphere. 

Because of the spotted structure of Ap stars, average abundances do not reflect 
the complexity of the physical conditions at the stellar surface, yet they are 
useful for statistical analysis and comparison of stars in the class (e.g., 
\citealt{Adelman73}; \citealt{Ryabchikova04}). HD\,75049 has a rich spectrum, but 
is not extraordinarily peculiar. In the spectrum many lines of rare earth elements 
are visible, but the majority of the strongest metal lines belong to 
Si\,\textsc{ii}, Ti\,\textsc{ii}, Fe\,\textsc{ii} and Cr\,\textsc{ii}. Lines of 
Sr\,\textsc{ii} in the blue region of the FEROS spectrum are also rather strong. 

{We obtained the abundances of several elements using spectral synthesis
where the observed spectra were compared with synthetic spectra until a 
best fit was obtained. An example for Fe\,\textsc{ii} $\lambda$5018\,\AA\ 
is shown in Fig.\,5.  The synthetic spectra were calculated with the software 
\textsc{SYNTHMAG} \citep{Piskunov99}. }
 The spectral line list was taken from the 
Vienna Atomic Line Database (VALD, \citealt{kupkaetal99}), which includes lines of 
rare earth elements from the DREAM database \citep{biemontetal99}. A model 
atmosphere with $T_{\rm eff} = 9600$\,K and $\log g = 4.0$ with metallicity of 
0.5\,dex above solar from the NEMO database \citep{heiteretal02} was used. 
The synthetic spectra provided a good match to many spectral lines with 
the field strength chosen, but some lines need further broadening to match the 
wings. Macroturbulence is possibly not the explanation for this, as magnetic 
fields suppress macroscopic motion in the stellar atmospheres. With the 
non-uniform distribution of abundances -- both horizontally and vertically 
-- in Ap 
star atmospheres, it may be that stronger magnetic field strengths in different 
line-forming regions could explain this extra broadening. Further study is needed 
to test this idea. 

Two spectra at rotation phases near minimum and maximum longitudinal field 
strength (phase 0.063 and 0.388; see Fig.\,\ref{fig:bzbm}) were used for 
abundances determination. Magnetic field strengths of 26\,kG and 30\,kG, 
respectively, were used for calculation of the synthetic spectra at these two 
rotational phases. Synthetic line profiles for various abundances were compared 
with observed profiles for best fits. Many calculated line profiles do not fit the 
observed profiles perfectly as a result of blending and asymmetry, but the average 
value for a number of lines gives reliable results. It was more difficult to fit 
the spectrum observed at rotational phase 0.063 when the field strength is lower 
and some lines are not as sharp and more asymmetric. Table\,4 presents the values 
of the abundances with errors determined for various ions. The errors given are 
the internal standard deviations for the list of examined lines for each ion. 
There were no lines of Nd\,\textsc{ii} found in the spectra, so only an upper 
limit is given in Table\,4. The last column in the table gives solar abundances 
for the same elements taken from \citet{asplundetal05}. HD\,75049 shows 
overabundances of silicon, chromium and europium and thus can be considered to 
belong to the SiCr group of Ap stars, according to the classification of peculiar 
stars. 

\begin{figure} 
\begin{center} 
\hfil \epsfxsize 8.5cm\epsfbox{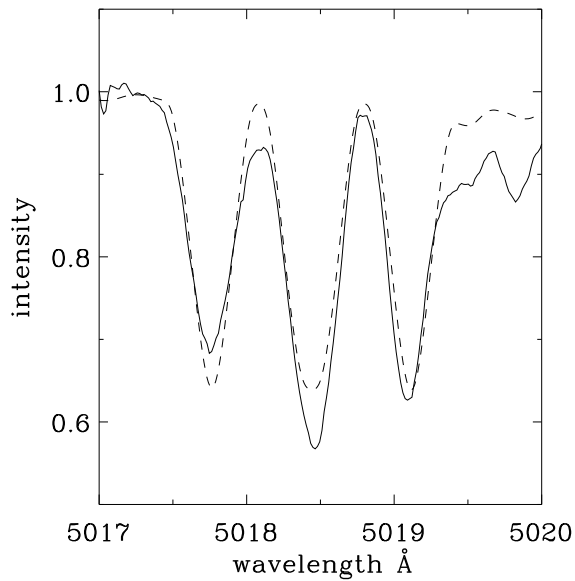} 
\caption{\label{fig:spm1} Observed (solid line) and synthetic (dashed line) 
profiles for Fe\,\textsc{ii} 5018\,\AA.}
\end{center}
\end{figure} 

\section{Concluding remarks} 

The strong magnetic field of HD\,75049 is an exciting discovery. This star is a 
rival for the long-known strongest Ap magnetic field in Babcock's star, 
HD\,215441. Both stars have deep silicon lines, but are at opposite ends of the 
temperatures range of the Ap\,Si stellar group. \citet{babcock60} noted that 
HD\,215441 does not have outstanding peculiarities in the spectrum. We have found 
the same in HD\,75049, which has a rich peculiar spectrum, but with overabundances 
of number chemical elements similar to other Ap stars with much smaller field 
strengths. 

HD\,215441 has a magnetic field modulus that changes from 32 to 35\,kG over the 
rotational period (\citealt{babcock60}; \citealt{Preston69}). \citet{Preston69} 
suggested that the magnetic field of this star deviates from a centred dipole. 
Magnetic configurations with a combination of dipole and quadrupole components 
\citep{Borra78} or dipole, quadrupole and octupole \citep{Landstreet00} components 
were required to model both observations of $B_z$ obtained by \citet{Borra78} and 
$\langle B \rangle$ by \citet{Preston69}.

{In contrast, for HD\,75049 the centred dipole model is a suitable first 
approximation to fit both the longitudinal field and field modulus measurements. 
Some discrepancy with the second magnetic moment dipole model may be a result of 
nonuniform distribution of chemical elements at surface. More complex magnetic 
geometry may also be present in this star. An important result is that the star is 
probably young. Its estimated stellar radius places it close to the zeor-age main 
sequence. }
       
\citet{Landstreet00} examined the inclination of rotation axis $i$ to 
line-of-sight and obliquity $\beta$ of the magnetic axis to the rotation 
axis for a sample 
of magnetic Ap stars. They found that the majority of the slow rotators with 
periods longer than 25\,d have $\beta$ smaller than $20^\circ$, while stars with 
short rotation periods tend to show a large obliquity of the magnetic axis. Our 
estimation of $\beta = 60^\circ \pm 3^\circ$ for HD\,75049 is consistent with 
this conclusion. 

\section{Acknowledgements} 
We thank Dr. Romanyuk for providing information about the Zeeman configuration of 
some spectral lines and for useful discussion. DWK and VGE acknowledge support for 
this work from the Science and Technology Facilities Council (STFC) of the UK. 
{This research has made use of NASA's Astrophysics Data System and SIMBAD 
database, operated at CDS, Strasbourg, France.}

{} 


\begin{thebibliography}{} 

\bibitem[\protect\citeauthoryear{Abt}{2009}]{Abt09} Abt H.~A., 2009, 
AJ, 138, 28 

\bibitem[\protect\citeauthoryear{Adelman}{1973}]{Adelman73} Adelman 
S.~J., 1973, ApJ, 183, 95 

\bibitem[\protect\citeauthoryear{Asplund, Grevesse, \& 
Sauval}{2005}]{asplundetal05} Asplund M., Grevesse N., Sauval 
A.~J., 2005, ASPC, 336, 25 

\bibitem[\protect\citeauthoryear{Babcock}{1947}]{babcock47} Babcock 
H.~W., 1947, ApJ, 105, 105 

\bibitem[\protect\citeauthoryear{Babcock}{1958}]{babcock58} Babcock 
H.~W., 1958, ApJS, 3, 141 

\bibitem[\protect\citeauthoryear{Babcock}{1960}]{babcock60} Babcock 
H.~W., 1960, ApJ, 132, 521 

\bibitem[\protect\citeauthoryear{Babel \& Lanz}{1992}]{Babel92} Babel 
J., Lanz T., 1992, A\&A, 263, 232 

\bibitem[\protect\citeauthoryear{Bi{\'e}mont, Palmeri, \& 
Quinet}{1999}]{biemontetal99} Bi{\'e}mont E., Palmeri P., Quinet 
P., 1999, Ap\&SS, 269, 635 

\bibitem[\protect\citeauthoryear{Borra \& Landstreet}{1978}]{Borra78} 
Borra E.~F., Landstreet J.~D., 1978, ApJ, 222, 226 

\bibitem[\protect\citeauthoryear{Cramer 
\& Maeder}{1979}]{Cramer79} Cramer N., Maeder A., 1979, A\&A, 78, 305 

\bibitem[\protect\citeauthoryear{Elkin et al.}{2008}]{Elkin08} Elkin 
V.~G., Kurtz D.~W., Freyhammer L.~M., Hubrig S., Mathys G., 2008, 
MNRAS, 390, 1250 

\bibitem[\protect\citeauthoryear{Freyhammer et 
al.}{2008}]{freyhammer08} Freyhammer L.~M., Elkin V.~G., Kurtz 
D.~W., Mathys G., Martinez P., 2008, MNRAS, 389, 441 

\bibitem[\protect\citeauthoryear{Freyhammer et 
al.}{2009}]{Freyhammer09} Freyhammer L.~M., Kurtz D.~W., Elkin 
V.~G., Mathys G., Savanov I., Zima W., Shibahashi H., Sekiguchi K., 
2009, MNRAS, 396, 325 

\bibitem[\protect\citeauthoryear{Harmanec}{1988}]{Harmanec88} Harmanec 
P., 1988, BAICz, 39, 329 

\bibitem[\protect\citeauthoryear{Hauck 
\& North}{1993}]{Hauck93} Hauck B., North P., 1993, A\&A, 269, 403 

\bibitem[\protect\citeauthoryear{Heiter et al.}{2002}]{heiteretal02} 
Heiter U., et al., 2002, A\&A, 392, 619 

\bibitem[\protect\citeauthoryear{Hubrig et al.}{2005}]{hubrigetal05} 
Hubrig S., et al., 2005, A\&A, 440, L37 

\bibitem[\protect\citeauthoryear{Hubrig et al.}{2009}]{hubrigetal09} 
Hubrig S., Mathys G., Kurtz D.~W., Sch{\"o}ller M., Elkin V.~G., 
Henrichs H.~F., 2009, MNRAS, 657 

\bibitem[\protect\citeauthoryear{Hubrig, North, \& 
Mathys}{2000}]{Hubrig00} Hubrig S., North P., Mathys G., 2000, 
ApJ, 539, 352 

\bibitem[\protect\citeauthoryear{Hubrig et al.}{2006}]{hubrigetal06} 
Hubrig S., North P., Sch{\"o}ller M., Mathys G., 2006, AN, 327, 289 

\bibitem[\protect\citeauthoryear{Kochukhov}{2006}]{Kochukhov06} 
Kochukhov O., 2006, A\&A, 454, 321 

\bibitem[\protect\citeauthoryear{Kochukhov et al.}{2004}]{Kochukhov04} 
Kochukhov O., Drake N.~A., Piskunov N., de la Reza R., 2004, A\&A, 
424, 935 

\bibitem[Kochukhov et al.(2009)]{koch09} Kochukhov, O., Shulyak, D., \& 
Ryabchikova, T.\ 2009, A\&A, 499, 851

\bibitem[\protect\citeauthoryear{Kudryavtsev et 
al.}{2006}]{kudretal06} Kudryavtsev D.~O., Romanyuk I.~I., Elkin 
V.~G., Paunzen E., 2006, MNRAS, 372, 1804 

\bibitem[\protect\citeauthoryear{Kupka et al.}{1999}]{kupkaetal99} 
Kupka F., Piskunov N., Ryabchikova T.~A., Stempels H.~C., Weiss 
W.~W., 1999, A\&AS, 138, 119 

\bibitem[\protect\citeauthoryear{Kurtz}{1985}]{Kurtz85} Kurtz D.~W., 
1985, MNRAS, 213, 773 

\bibitem[\protect\citeauthoryear{Kurtz et al.}{2006}]{kurtzetal06} 
Kurtz D.~W., Elkin V.~G., Cunha M.~S., Mathys G., Hubrig S., Wolff 
B., Savanov I., 2006, MNRAS, 372, 286 

\bibitem[\protect\citeauthoryear{Kurucz}{1979}]{kurucz79} Kurucz 
R.~L., 1979, ApJS, 40, 1 

\bibitem[\protect\citeauthoryear{Landi Degl'Innocenti}{1975}]{landi75} 
Landi Degl'Innocenti, E., 1975, A\&A, 45, 269 

\bibitem[\protect\citeauthoryear{Landstreet \& 
Mathys}{2000}]{Landstreet00} Landstreet J.~D., Mathys G., 2000, 
A\&A, 359, 213 

\bibitem[\protect\citeauthoryear{Lenz \& Breger}{2005}]{lenz05} Lenz 
P., Breger M., 2005, CoAst, 146, 53 

\bibitem[\protect\citeauthoryear{Martinez}{1993}]{mart93} Martinez P., 
1993, PhD Thesis, University of Cape Town 

\bibitem[\protect\citeauthoryear{Mathys}{1990}]{Mathys90} Mathys G., 
1990, A\&A, 232, 151 

\bibitem[\protect\citeauthoryear{Mathys}{1991}]{Mathys91} Mathys G., 
1991, A\&AS, 89, 121 

\bibitem[\protect\citeauthoryear{Mathys}{1995}]{Mathys95} Mathys G., 
1995, A\&A, 293, 746 

\bibitem[\protect\citeauthoryear{Mathys et al.}{1997}]{Mathys97} 
Mathys G., Hubrig S., Landstreet J.~D., Lanz T., Manfroid J., 1997, 
A\&AS, 123, 353 

\bibitem[\protect\citeauthoryear{Mathys et al.}{2000}]{Mathys00} 
Mathys G., Stehl\'e C., Brillant S., Lanz T. 2000, A\&A, 358, 1151 

\bibitem[\protect\citeauthoryear{Mermilliod, Mermilliod, 
\& Hauck}{1997}]{Mermilliod97} Mermilliod J.-C., Mermilliod M., Hauck B., 1997, 
A\&AS, 124, 349 

\bibitem[\protect\citeauthoryear{Moon \& Dworetsky}{1985}]{Moon85} 
Moon T.~T., Dworetsky M.~M., 1985, MNRAS, 217, 305 


\bibitem[\protect\citeauthoryear{Munari 
\& Zwitter}{1997}]{Munari97} Munari U., Zwitter T., 1997, A\&A, 318, 269 

\bibitem[Napiwotzki et al.(1993)]{napi93} Napiwotzki, R., Schoenberner, D., \& 
Wenske, V.\ 1993, A\&A, 268, 653 

\bibitem[\protect\citeauthoryear{North \& Kroll}{1989}]{North89} North 
P., Kroll R., 1989, A\&AS, 78, 325 

\bibitem[\protect\citeauthoryear{North 
\& Nicolet}{1990}]{North90} North P., Nicolet B., 1990, A\&A, 228, 78 

\bibitem[\protect\citeauthoryear{Pojmanski}{2002}]{Pojmanski02} 
Pojmanski G., 2002, AcA, 52, 397 

\bibitem[\protect\citeauthoryear{Piskunov}{1992}]{piskunov92} 
Piskunov, N.~E.\ 1992, Stellar Magnetism, 92 

\bibitem[\protect\citeauthoryear{Piskunov}{1999}]{Piskunov99} Piskunov 
N., 1999, ASSL, 243, 515 

\bibitem[\protect\citeauthoryear{Preston}{1967}]{Preston67} Preston 
G.~W., 1967, ApJ, 150, 547 

\bibitem[\protect\citeauthoryear{Preston}{1969}]{Preston69} Preston 
G.~W., 1969, ApJ, 156, 967 

\bibitem[\protect\citeauthoryear{Preston}{1970}]{Preston70} 
Preston G.~W., 1970, ApJ, 160, 1059 

\bibitem[\protect\citeauthoryear{Ryabchikova et 
al.}{2004}]{Ryabchikova04} Ryabchikova T., Nesvacil N., Weiss 
W.~W., Kochukhov O., St{\"u}tz C., 2004, A\&A, 423, 705 

1992, 

\bibitem[\protect\citeauthoryear{Stibbs}{1950}]{Stibbs50} Stibbs 
D.~W.~N., 1950, MNRAS, 110, 395 

\bibitem[\protect\citeauthoryear{Valyavin et al.}{2007}]{Valyavin07} 
Valyavin G., Lee B.-C., Shulyak D., Han I., Kochukhov O., Khang 
D.-I., Kim K.-M., 2007, ASPC, 362, 245 

\bibitem[\protect\citeauthoryear{Wade et al.}{2001}]{Wade01} Wade 
G.~A., Ryabchikova T.~A., Bagnulo S., Piskunov N., 2001, ASPC, 248, 
373 

\end{thebibliography}
\end{document}